\documentclass[conference]{IEEEtran}
\usepackage{url}
\usepackage{graphicx}
\usepackage{amsfonts}
\usepackage{amsmath, amssymb, bm}
\usepackage[]{algorithm}
\usepackage[]{algpseudocode}
\usepackage{listings}
\usepackage{cite}
\usepackage{tikz}
\usepackage{subcaption}
\usepackage{multicol}

\usetikzlibrary{fit,positioning, calc}

\newcommand{\ds}{\displaystyle}

\newcommand{\R}{\mathbb{R}}
\usepackage{color}

\newcommand{\cond}{\; | \;}

\DeclareMathOperator*{\argmax}{argmax}

\begin{document}
%

\title{Towards Generic Deobfuscation\\ of Windows API Calls}



\author{\IEEEauthorblockN{Vadim Kotov}
\IEEEauthorblockA{Dept. of Research and Intelligence\\Cylance, Inc\\vkotov@cylance.com}
\and
\IEEEauthorblockN{Michael Wojnowicz}
\IEEEauthorblockA{Dept. of Research and Intelligence\\Cylance, Inc\\mwojnowicz@cylance.com}}

\IEEEoverridecommandlockouts
\makeatletter\def\@IEEEpubidpullup{9\baselineskip}\makeatother
\IEEEpubid{\parbox{\columnwidth}{
    Workshop on Binary Analysis Research (BAR) 2018 \\
    18 February 2018, San Diego, CA, USA\\
    ISBN 1-891562-50-9\\
    https://dx.doi.org/10.14722/bar.2018.23011\\
    www.ndss-symposium.org
}
\hspace{\columnsep}\makebox[\columnwidth]{}}

%

\maketitle

\begin{abstract}
  A common way to get insight into a malicious program's functionality is to look at which API functions it calls.  To complicate the reverse engineering of their programs, malware authors deploy API obfuscation techniques, hiding them from analysts' eyes and anti-malware scanners. This problem can be partially addressed by using dynamic analysis; that is, by executing a malware sample in a controlled environment and logging the API calls. 
However, malware that is aware of virtual machines and sandboxes might terminate without showing any signs of malicious behavior. In this paper, we introduce a static analysis technique allowing generic deobfuscation of Windows API calls. The technique utilizes symbolic execution and hidden Markov models to predict API names from the arguments passed to the API functions. Our best prediction model can correctly identify API names with  87.60\% accuracy.
\end{abstract}

\section{Introduction}
\label{sec:introduction}

Malware plays by the same rules as legitimate software, so in order to do something meaningful (read files, update the registry, connect to a remote server, etc.) it must interact with the operating system via the Application Programming Interface (API). On Windows machines, the API functions reside in dynamic link libraries (DLL). Windows executables \cite{msdn-pe-format} store the addresses of the API functions they depend on in the Import Address Table (IAT) - an array of pointers to the functions in their corresponding DLLs. Normally these addresses are resolved by the loader upon program execution.

When analyzing malware, it is crucial to know what API functions it calls - this provides good insight into its capabilities \cite{arntz-2017, gennari-2017}. That is why malware developers try to complicate the analysis by obfuscating the API calls \cite{symantec-api-obf-museum}.  When API calls are obfuscated, the IAT is either empty or populated by pointers to functions unrelated to malware's objectives, while the true API functions are resolved on-the-fly.  This is usually done by locating a DLL in the memory and looking up the target function in its Export Table - a data structure that describes API functions exposed by the DLL.  In other words, obfuscated API calls assume some ad-hoc API resolution procedure, different from the Windows loader.

Deobfuscating API calls can be tackled in two broad ways:

\begin{enumerate}
\item Using static analysis, which requires reverse engineering the obfuscation scheme and writing a script that puts back missing API names.
\item Using dynamic analysis, which assumes executing malware in the controlled environment and logging the API calls.
\end{enumerate}

Static analysis allows exploration of every possible execution branch in a program and fully understand its functionality. Its major drawback is that it can get time consuming as some malware families deploy lengthy and convoluted obfuscation routines (e.g. Dridex banking Trojan \cite{su-2015}). Furthermore, even minor changes to the obfuscation schemes break the deobfuscation scripts, forcing analysts to spend time adapting them or re-writing them altogether. Dynamic analysis, on the other hand, is agnostic of obfuscation, but it can only explore one control flow path, making the analysis incomplete. Additionally, since dynamic analysis is usually performed inside virtual machines (VM) and sandboxes, a VM-/sandbox-aware malware can potentially thwart it.

In this paper, we introduce a static analysis approach, allowing generic deobfuscation of Windows API calls. Our approach is based on an observation that malware analysts can often ``guess'' some API functions by just looking at their arguments and the context in which they are called. For example, consider \texttt{RegCreateKeyEx}:

\begin{small}
\begin{verbatim}
LONG WINAPI RegCreateKeyEx(
1.  HKEY                  hKey,
2.  LPCTSTR               lpSubKey,
3.  DWORD                 Reserved,
4.  LPTSTR                lpClass,
5.  DWORD                 dwOptions,
6.  REGSAM                samDesired,
7.  LPSECURITY_ATTRIBUTES lpSecurityAttributes,
8.  PHKEY                 phkResult,
9.  LPDWORD               lpdwDisposition
);
\end{verbatim}
\end{small}

Arguments 5, 6, 7 and 9 are pre-defined constants (permission flags, attributes etc.) and can only take a finite and small number of potential values (it's also partially true for the argument 1, which aside from a set of pre-defined values can take an arbitrary one). So if we see $\geq 9$ arguments on the stack and values 1, 5, 6, 7, 9 are registry related constants, we can conclude that \texttt{RegCreateKeyEx} is being called even if we don't see the actual name.  Similar arguments can be made about \texttt{CreateProcess}, \texttt{VirtuaAlloc}, etc.

Our goal is to automate this inferential process and see whether WinAPI functions' names can be predicted by the arguments passed to them upon calling.  This is the kind of problem where machine learning shows promising results, similar in nature to image or signal recognition \cite{rabiner-1990, carbonell-1983}.

In this work, we evaluate the feasibility of the proposed idea.  To do so, we first solve a simplified problem:
\begin{itemize}
\item We look only at functions with 3 or more arguments.
\item We limit our dataset to standard 32-bit Windows executables and DLLs (e.g. Internet Explorer, cmd.exe, etc.).
\item We focus on the 25 most used API functions.
\item We ignore any adversarial techniques such as control-flow obfuscation, or anti-disassembly tricks.
\end{itemize}

We extract API calls (function names and their arguments) from our dataset of Windows binaries and use them to train a machine learning pipeline which uses Hidden Markov Models (HMM) for vectorization and Multinomial Logistic Regression (MLR) for prediction. 

To extract the dataset of API calls, we build a simplified symbolic execution engine. We call it ``simplified" because it supports only a small subset of x86 instructions, just enough to keep track of the arguments passed to API functions.

We conducted two experiments:

\begin{itemize}
\item In the first experiment, we test the feasibility of API prediction by arguments and whether ordering of the arguments matters. For this experiment we assume the number of arguments of an API function being predicted is known.
\item In the second experiment, we create a more realistic scenario where the number of arguments is unknown. We  build a refined model and test if it can (implicitly) infer the number of arguments indented for an API call.
\end{itemize}
  
The results show that our models can predict API names by the arguments with the accuracy of 73.18\% in the first experiment (when the number of arguments is known) and 87.60\% in the second one (when the number of the arguments is unknown, but the methodology is improved).

\subsection{Source Code}
The source code of the tools developed for this research as well as instructions on how to use them can be found at \url{https://github.com/cylance/winapi-deobfuscation}. The particular tagged commit is at \url{https://github.com/cylance/winapi-deobfuscation/releases/tag/bar-2018}.


\section{Related Work}
\label{sec:related}
  
One of the first works to address API obfuscation problems was the Eureka framework \cite{sharif-2008}. This is a generic malware analysis framework and API deobfuscation is only one side of it. Eureka's API deobfuscation workflow includes the following steps:

\begin{enumerate}
\item \emph{Building look up tables} - Eureka builds up a look up table of the API functions exported by the DLLs that are loaded at pre-defined addresses (this assumes the ASLR\footnote{ASLR stands for address space layout randomization - an exploitation mitigation which enables the system to load executables at random address locations} to be disabled). For the DLLs loaded at randomized addresses (e.g. when ASLR is enabled) it watches out for the DLLs being loaded by the malicious program.
\item \emph{Identifying candidate API calls} - it statically identifies potential API calls by flagging call targets outside the program's inter-procedural control flow graph.
\item \emph{Identifying static call addresses} -  for the flagged candidate calls it looks up the call targets in the pre-built API tables. This method is aimed at a particular obfuscation technique, where API functions are called directly by their addresses in the DLLs bypassing the IAT.
\item \emph{Identifying dynamically computed call addresses} - it monitors calls to \texttt{GetProcAddress}\footnote{An API function resolving other API functions by their names} and memory write operations. When a memory write is encountered it checks whether \texttt{GetProcAddress} has been called earlier. If \texttt{GetProcAddress} is found in the control flow path, then the API name being resolved is \texttt{GetProcAddresses}'s second argument.
\end{enumerate}

The approach used in Eureka has a strong focus on call targets, which leads to the following limitations: (1) the static analysis method (Step 3) can be bypassed by obfuscating the target address using a technique Eureka is unaware of; (2) the dynamic analysis method (Step 4) can be bypassed by locating the address of an API in the memory space of a DLL without using GetProcAddress.

The authors of \cite{xi-2013} focus on API obfuscation schemes which encrypt the API names. In such a scheme a decryptor first obtains the name of an API function and then uses \texttt{GetProcAddress} to resolve its address. The authors devised a decryptor-agnostic approach to extract the names of imported API functions. It deploys program slicing and taint propagation to correlate a potential decryption routine with the call to \texttt{GetProcAddress}. Once the API decryptor is identified they use a debugger as a binary instrumentation tool to ``emulate'' the decryption and extract the true API function address. This approach suffers from the same problem as the previous one - it uses \texttt{GetProcAddress} as a means of finding the API decryptor.

The approach of \cite{choi-2015} is better in that it doesn't rely on \texttt{GetProcAddress}. The author makes the distinction between static (or compile-time) and dynamic (or run-time) obfuscation. In static obfuscation the API resolution procedure is hardcoded and stays the same from execution to execution. Meanwhile, dynamic obfuscation assumes copying and obfuscating the code from the DLLs into a newly allocated executable memory area and then jumping into that area. The author proposes two deobfuscation methods - one for the dynamic obfuscation and one for the static one:

\begin{itemize}
\item To defeat dynamic obfuscation, the author proposes using memory access analysis that correlates memory read operations from the executable areas of DLLs and subsequent memory write operations into newly allocated memory regions. 
\item A method for static obfuscation checks whether a call target points at an address of an API function and if it does, the obfuscated call is replaced with the address of the API function.
\end{itemize}

The limitation of this approach and dynamic analysis methods in general is that sometimes dynamic analysis is impractical or impossible. For example, an analyst might be given a corrupted executable image that won't run or a malware sample that deploys anti-VM/anti-debugging techniques.

Ours is a static analysis approach and therefore doesn't require execution, or a malware sample to be complete. Static analysis is a challenging discipline as even simple obfuscation might thwart static analyzers. Our contribution does not eliminate the need for dynamic analysis, but improves the feasibility of static analysis.

\section{Background}
\label{sec:background}

\subsection{Symbolic Execution}
Symbolic execution \cite{king-1976, baldoni-2016} is a type of program execution that can operate on symbolic values. It can be seen as an extension of concrete execution where unknown values (e.g. program's user input) are represented by symbols. Computation over symbolic values then is essentially building up a formula. For example, consider the x86 instruction \texttt{add eax, 5}. It can't be executed concretely as the value of \texttt{eax} is unknown. But in symbolic execution the result will be a symbolic formula \texttt{eax + 5}.\footnote{For simplicity we ignore side effects in this example.}

\begin{figure}
  \includegraphics[width=0.5\textwidth]{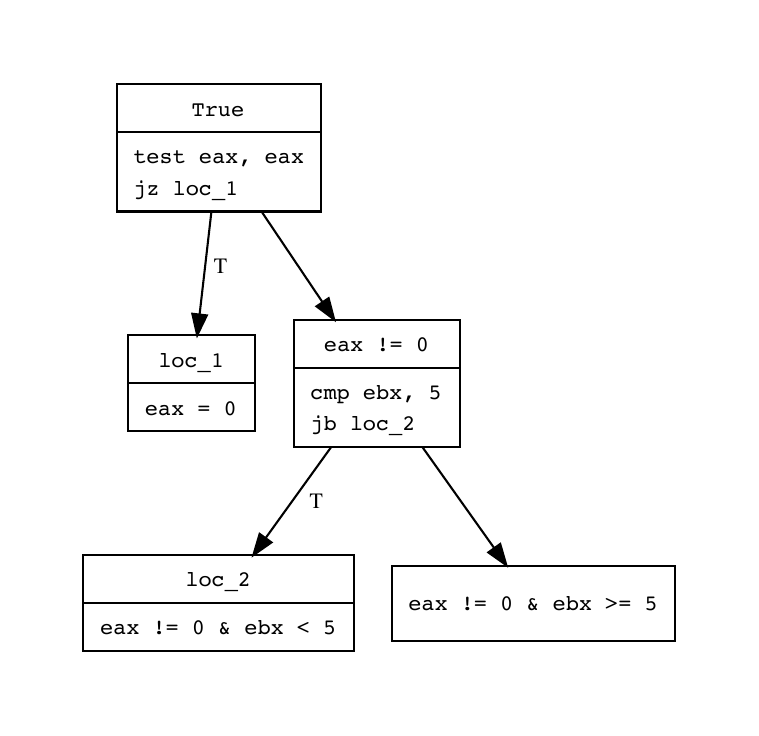}
  \caption{Illustration of the constraints for each control flow path. At first, the formula is set to simply ``True'' as no constraints exist yet. As the program branches, a set of constraints gets bigger and bigger for each branch. Solving for the constraints defines what input values will result into program taking that path.}
  \label{img:symex-example}
\end{figure}

Another feature of symbolic execution is the ability to examine all the control flow paths of a program. For each path there is a first-order logic (FOL) formula that describes what conditions must be satisfied for the program to take that path (see Figure \ref{img:symex-example}). These formulas then can be checked with a satisfiability solver to see if a path is realizable or find out what inputs are needed for best code coverage etc.

The ability to execute every path comes at a price - applying symbolic execution to large functions and programs might get prohibitively expensive - a problem known as path explosion. In practice only a subset of all paths is examined based on problem-specific criteria.

A symbolic execution engine should be able to model memory. The most straightforward model is, perhaps, fully symbolic memory, the simplest example of which is a key-value storage. A good overview of memory modeling strategies as well as other aspects of symbolic execution can be found in \cite{baldoni-2016}.

Originally created to improve code coverage for software testing \cite{king-1976}, symbolic execution has applications in security, e.g. malware analysis assistance \cite{baldoni-2017}, identifying trigger behavior in malware \cite{brumley-2008}, augmenting fuzzing techniques \cite{stephens-2016}, checking stability of patches and identifying bugs \cite{ramos-2015}, etc.

In our work, however, we use a simplified version of symbolic execution, which: 

\begin{itemize}
\item can only execute a small subset of x86 instructions;
\item does not support branching;
\item does not support function calls.
\end{itemize}

These limitations were introduced for the sake of performance efficiency and to avoid path explosion as we wanted to generate the dataset for experimentation as fast as possible. At this stage of the research we are less interested in code coverage and more in generating the training dataset for our machine learning models. Our simplified symbolic execution engine is described in Section \ref{sec:sym-ex}. In the future research we are planning to use a full-functioning symbolic execution engine.

\subsection{Hidden Markov Models}

Hidden Markov models (HMMs)~\cite{rabiner1986introduction},~\cite{beal2003variational} are a popular statistical model for sequential data,\footnote{A deep learning alternative would be Long Short Term Memory (LSTM) neural networks; however, we stick with the simpler HMM model for the proof of concept, especially because our dataset is still relatively small, and plausibly within the regime where HMMs outperform LSTMs~\cite{panzner2016comparing}.} finding success in diverse fields such as speech recognition, computational molecular biology,  data compression, cryptanalysis, and finance.  In particular, as the field of cybersecurity begins to incorporate data mining and machine learning methods~\cite{dua2016data, wojnowicz2017suspend, silva2017improving, wojnowicz2016influence}, HMMs have become increasingly popular tools in cybersecurity; for instance, they have been used to detect intrusions in substations of a power system~\cite{ten2011anomaly}, to learn user profiles of web browsing behavior~\cite{abramson2014learning}, and to detect anomalous programs based on system call patterns~\cite{wang2006profiling}.

\begin{figure*}[h]
  \centering
  \begin{subfigure}{.46\textwidth}
  
  \includegraphics[width=\linewidth]{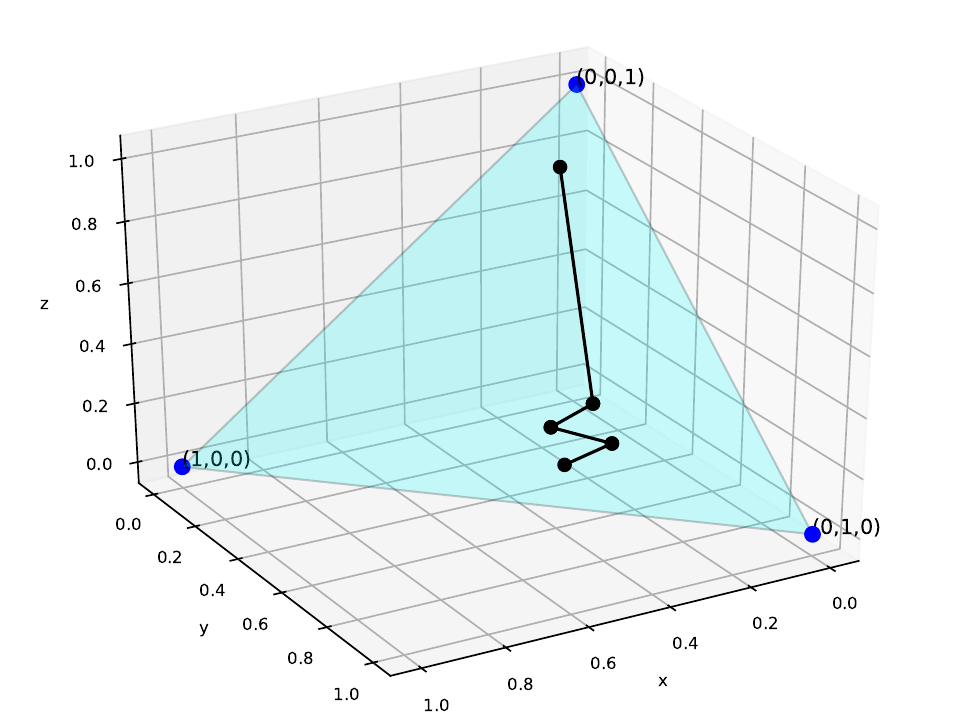}
\caption{A latent trajectory on a standard 2-simplex in $\R^3$.}
  \label{simplex}
\end{subfigure}
 \begin{subfigure}{.46\textwidth}
\centering
\resizebox {\columnwidth}{!}{
\begin{tikzpicture}
\tikzstyle{main}=[circle, minimum size = 10mm, thick, draw =black!80, node distance = 16mm]
\tikzstyle{connect}=[-latex, thick]
\tikzstyle{box}=[rectangle, draw=black!100]
  \node[main, fill = white!100] (y1) [label=below: \large $y_1$] { };
    \node[main, fill = white!100] (s1) [above=of y1, label=above:\large  $s_1$] { };
     \node[main, fill = white!100] (y2) [right=of y1,label=below:\large$y_2$] { };
    \node[main, fill = white!100] (s2) [above=of y2, label=above:\large$s_2$] { };
        \node[main, fill = white!100] (y3) [right=of y2,label=below:\large$y_3$] { };
    \node[main, fill = white!100] (s3) [above=of y3, label=above:\large$s_3$] { };
          \node[main, fill = white!100] (yT-1) [right=of y3,label=below:\large$y_{T-1}$] { };
    \node[main, fill = white!100] (sT-1) [above=of yT-1, label=above:\large$s_{T-1}$] { };
       \node[main, fill = white!100] (yT) [right=of yT-1,label=below:\large$y_T$] { };
    \node[main, fill = white!100] (sT) [above=of yT, label=above:\large$s_T$] { };
    
     \node at ($(s3)!.5!(sT-1)$) {\ldots}; 
      \node at ($(y3)!.5!(yT-1)$) {\ldots}; 
 \path (s1) edge [connect] (y1); 
 \path (s2) edge [connect] (y2);
  \path (s3) edge [connect] (y3);
     \path (sT-1) edge [connect] (yT-1);
   \path (sT) edge [connect] (yT);
 \path (s1) edge [connect] (s2)
 	 (s2) edge [connect] (s3);
 \path (sT-1) edge [connect] (sT);
\end{tikzpicture}
}
\caption{Graphical model representation of a HMM.}
\label{graphical_model_HMM}
\end{subfigure}
\caption{Conceptual overview of a Hidden Markov Model (HMM)}
\end{figure*}
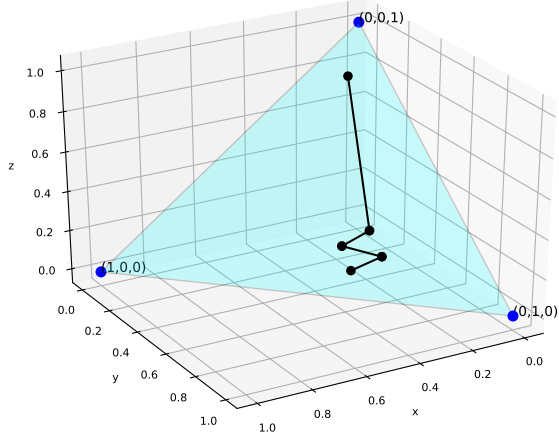
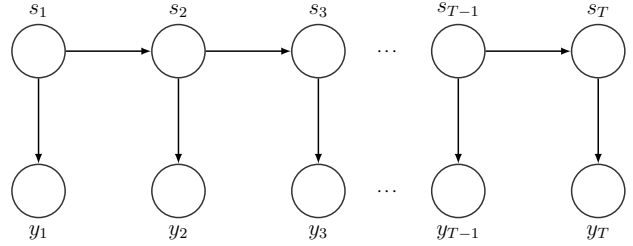

HMMs are generalizations of mixture models (MMs).  
To build intuition, consider the standard n-simplex, defined by:
\[ \Delta^n = \bigg\{  (t_0, \hdots, t_n) \in \R^{n+1} \cond \ds\sum_{i=0}^n t_i =1 \; \text{and} \; t_i \geq 0 \; \text{for all} \; i \bigg\} \]
(See Fig.~\ref{simplex}.) MMs represent an estimated probability distribution as a point on the simplex.    Thus, for MM's, the vertices represent probability distributions (the ``mixture components", typically summarized by fitted parameters from a parametric family, such as means $\mu_i$ and covariance matrices $\Sigma_i$ from Gaussian distributions), and the location of a point on the simplex represents the mixture weights.    HMMs  are like MMs in that the vertices represent distributions.   However, whereas MMs describe an entire dataset as a single point on the simplex, HMMs assign mixture weights to  \emph{each} observation, such that the observed sequence is modeled as traveling through mixture weight space.

For HMMs to map \emph{individual} observations onto the simplex, they must make additional assumptions about sequence dynamics which do not exist for models, such as MMs, which assume each data point is independent and identically distributed.  
To state them, let $y_{1:T} = \{y_1, \hdots y_T \}$ be a sequence of observations, and $s_{1:T} = \{s_1, \hdots s_T \}$ be a corresponding sequence of unobserved hidden states, where each $s_t$ is a discrete K-valued random variable (i.e. each $s_t \in \Delta^{K-1}$).   Then the assumptions for HMM are:
\begin{enumerate}
\item \emph{(First order) Markovian hidden state transitions:} The $t$th hidden state is conditionally independent of previous data except for the $(t-1)$st hidden state.
\[P(s_t \cond s_{1:t-1}, y_{1:t-1}) = P(s_t \cond s_{t-1})\]
\item \emph{Time homogeneity}:  The underlying ``hidden" Markov chain defined by $P(s_t \cond s_{t-1})$ is independent of time $t$ (i.e., is stationary). \label{homogeneity}
\item \emph{Conditionally Independent Observations:} The $t$th observation is conditionally independent of all variables except for the $t$th hidden variable
\[P(y_t \cond s_{1:T}, y_{1:T}) = P(y_t \cond s_t) \]
\end{enumerate}
Figure \ref{graphical_model_HMM} provides a graphical model representation, where edges represent dependencies.  The graphical model reflects the assumptions above:  the hidden states follow first-order Markov dynamics (edges between $s_t$ and $s_{t+1}$), and the dependence between observations is completely accounted for by the unobserved hidden state process (edges between $s_t$ and $y_t$).

Given these assumptions, the model is characterized by parameters $\lambda=(\pi, A, \theta)$:
\begin{align*}
\scriptsize \text{Initial state distribution} && \pi = \{\pi_i\}  &= P(s_1 = i) \\
\scriptsize \text{State transition matrix} && A=\{a_{i,j}\} &= P(s_t = j \cond s_{t-1}=i)  \\
\scriptsize \text{Emission distribution } && \theta = \{ \theta_j\}& \; \tiny\text{for}\normalsize \; P(y_t \cond s_t=j, \theta_j)   \\
\scriptsize \text{parameters} && 
\end{align*}
We can interpret the parameters with respect to  Figure \ref{graphical_model_HMM}. The initial state distribution, $\pi$,  describes the probability, before any data is observed, of various values of the initial hidden state $s_1$ (say $s_1=k$) in the top left node of the Figure.   The $k$th emissions distribution parameter, $\theta_k$, then determines the probability of the first observation, $y_1$, given that $s_1=k$. The state transition matrix $A$ describes the probability of transitioning from the first latent state, $s_1$, to each possible latent state for $s_2$.  Following this logic through the graphical model, we obtain the \emph{complete data likelihood} of a sequence of length $T$, which is (suppressing $\lambda$ for simplicity):
\begin{equation}
p(s_{1:T}, y_{1:T}) = p(s_1)p(y_1 \cond s_1) \ds\prod_{t=2}^T p(s_t \cond s_{t-1}) p(y_t \cond s_t) 
\label{complete_data_likelihood}
\end{equation}

A trained HMM model allows one to evaluate the likelihood of a sequence, $P(y_{1:T} \cond \lambda)$.   However, this evaluation is not immediate using (\ref{complete_data_likelihood}), because the hidden state process is unobserved.   One might attempt a naive marginalization via enumeration, i.e.
\begin{equation}
\label{marginalization}
P(y_{1:T} \cond \lambda)  = \ds\sum_{s_{1:T}}P(y_{1:T}, s_{1:T} \cond \lambda) 
\end{equation} 
However, note that this summation is over $K^T$ state sequences.   In contrast, a recursive \emph{forward algorithm} uses ideas from dynamic programming to achieve efficient $\mathcal{O}(TK^2)$ likelihood evaluations.   This algorithm exploits the Markovian structure of the model to simplify the number of computations of the likelihood evaluation. 

Although notationally suppressed, all probability factors in (\ref{complete_data_likelihood}) are conditioned on $\lambda$.  A maximum likelihood estimate of $\lambda$ can be obtained by employing the Baum-Welch algorithm (a special case of the well-known Expectation Maximization (EM) algorithm).   This is done by setting an initial guess for $\lambda$, and then iterating over the following steps until a stopping criterion is satisfied:
\begin{enumerate}
\item \emph{E-step}: Compute $\widehat{s}_{1:T} = \mathbb{E}[s_t = k | y_{1:T}, \lambda]$ for all $t, k$.\footnote{ The \emph{forward-backward} algorithm, derived from dynamic programming, makes the E-step computationally tractable.}
\item \emph{M-step:} Given $\widehat{s}_{1:T}, y_{1:T}$, find $\lambda$ maximizing (\ref{complete_data_likelihood}).
\end{enumerate}
This procedure is guaranteed to monotonically converge to a local maximum of the likelihood.    Moreover, it can easily be modified to handle multiple sequences (as occur in our dataset)~\cite{rabiner1986introduction}.  Simply perform the E-step for each sequence separately, and then perform the M-step where the complete data likelihood is adjusted into a product over multiple observed sequences. 

Fig.~\ref{simplex} allows for a geometric interpretation of Baum-Welch. The M-step corresponds to ``re-interpreting" the simplex (adjusting the following: the value of the emissions distribution parameters at each vertex, the starting point for the trajectories, and the degree of resistance to traveling in different directions), and the E-step corresponds to finding the best simplex trajectory (or trajectories) to match the observed dataset given the current simplex ``interpretation" provided by the M-step. 

For our purposes, the mapping of sequences onto the simplex produces a compact representation from which we can derive useful features for our predictive model. 

\section{The Proposed Approach}
\label{sec:approach}
\subsection{Simplified Symbolic Execution Engine Overview}
\label{sec:sym-ex}
The simplified symbolic execution engine is a research tool we developed to collect the training data. It is designed to process functions as opposed to full programs. It supports a limited number of x86 instructions that we assume to be crucial for API call recognition and ignores the rest. The list of supported instructions and explanations for why they were selected are shown in Table \ref{tab:supported-instructions}.

\begin{table}[h]
\centering
\caption{The subset x86 instructions supported by the simplified symbolic execution engine}
\label{tab:supported-instructions}
\begin{tabular}{| p{1.5cm} | p{6cm} |}
\hline
Instructions & Explanation\\
\hline \hline
\texttt{push, pop} &  Windows API functions follow standard calling convention (\_\_stdcall) \cite{msdn-winapi-stdcall}, i.e. the arguments are pushed to the stack in reverse order. \texttt{pop} must be supported alongside \texttt{push} to properly model the stack. \\
\hline
\texttt{mov} & Used to transfer data between the registers and between the registers and the memory. Sometimes used to put an address of an API function into a register  e.g. \texttt{mov esi, RegOpenKeyExW / ... / call esi}.\\
\hline
\texttt{lea} & Used to transfer memory addresses. E.g. \texttt{lea eax, [ebp+phkResult]} puts the address of the variable \texttt{phkResult} into \texttt{eax} (as opposed to its value)\\
\hline
\texttt{xor} & Often used to set a register value to zero. Many WinAPI arguments  are zeros or NULL-pointers.\\
\hline
\texttt{add, sub, inc, dec} & Sometimes addition and subtraction (including increment and decrement operations) are used instead of \texttt{mov}. E.g. \texttt{xor eax, eax / add eax, 25h / push eax} .\\
\hline
\end{tabular}
\end{table}

To speed up processing time and avoid path explosion, our engine executes the longest path through the function's control flow graph (CFG)\footnote{A CFG is a directed graph where vertices are basic blocks of the function and edges are control flows between them. A basic block is a sequence of instructions not interrupted by intra-procedural control flow transfer instructions.} ignoring loops.

We define a function CFG $G$ with $n$ vertices and $m$ edges as a tuple $(V_G, E_G, v_0, r)$, where
\begin{itemize}
\item $V_G = \{v_0, v_1, ..., v_n \}$ - is the set of vertices or nodes.
\item $E_G  \subset V_G \times V_G$ - is the set of $m$ edges.
\item $v_0 \in V_G$ - is the entry node - the point where the function takes the control flow.
\item $r \subseteq V_G$ - the set of return nodes, i.e. nodes from which the function returns or nodes with tail jumps\footnote{Tail jump is a compiler optimization where if a function $f_1$ calls a function $f_2$ right before returning (i.e. \texttt{call $f_2$ / ret}), the call is replaced with a jump to $f_2$. And then $f_2$ returns the control back to the caller.}.
\end{itemize}

A path through a function is a sequence of connected vertices starting at $v_0$ and ending at any $v_i \in r$. 
Let $P = \{p_0, p_1, ..., p_{|r|}\}$, be the set of all the function's paths, then the longest path $p_j$ is such that $|p_j| \geq |p_k|, \forall p_k \in P$, $j \neq k$.


Searching the longest path can be computationally expensive for functions whose CFG has a large number of edges. So if $|E_G| > 100$ we approximate the longest path by performing a random walk from $v_0$ to any $v_i \in r$ 30 times and picking the longest path out of 30. 

Once the longest path is identified the symbolic execution engines follows its basic blocks executing supported instructions. Upon reaching a \texttt{call} instruction it does one of the two things:
\begin{enumerate}
\item If the call address is not an API function (i.e. it's a call to another function in the program) - it ignores the call. 
\item If the call address is an API function or a function jumping to an API function (i.e. Delphi-style API call) - it reads at most 12 arguments from the stack (12 is the maximum possible number of arguments in the set of supported API calls); then it looks up the number of arguments it takes in the pre-constructed API database; and, finally, removes this many arguments from the stack (to keep the stack in consistent state). 
\end{enumerate}

In both cases it sets the \texttt{eax} register (where the returned value is supposed to be) to the dummy symbolic value ``ret''.

The symbolic execution engine is based on radare2 \cite{radare2} and uses its intermediate language called ESIL (Evaluable Strings Intermediate Language).

\subsection{Simulating Memory}
Our simulated memory is a key-value storage, where keys represent memory addresses written to or read from and values represent the data. Both memory addresses and data can be either symbolic or concrete.

Keys are strings of the format \texttt{size:addr} where
\begin{itemize}
\item \texttt{size} indicates the size of a memory operation. On a 32-bit x86 system it can be 1,2,4 bytes. 
\item \texttt{addr} is a string representation of the memory address.
\end{itemize}

Here's an example of memory write operation:
\begin{center}
  \begin{footnotesize}
    \texttt{mov [ebp-0xC], 0x1000} $\rightarrow$ \texttt{mem["4:ebp-0xC"] = 0x1000}
  \end{footnotesize}
\end{center}

Where \texttt{mem} is the key value storage. 

Memory read operations are simulated in a similar manner - we construct the key and fetch the value stored at that key. Sometimes the value of a particular address is unknown. In this case we emit a new symbolic value:
\begin{itemize}
\item $arg\_xh$, if the memory address is a function argument, i.e it is represented as \texttt{ebp+x}. E.g. a symbolic value for the memory address \texttt{ebp+0x8} is $arg\_8h$.
\item $var\_xh$, if the memory address is a local variable, i.e. it is represented as \texttt{ebp-x}. E.g. a symbolic value for the memory address \texttt{ebp-0xC} is $var\_Ch$.
\item $m_i$, where $i \in {0, 1, 2, ...}$ is a counter that increments each time a new value is emitted for any other address.
 \end{itemize}

Examples:

\begin{center}
  \begin{footnotesize}
    \texttt{mov esi, [ebp-0x8]} $\rightarrow$ \texttt{mem["4:ebp+0x8"] = arg\_8h} \\
    \texttt{mov cx, WORD [0x1068EEC]} $\rightarrow$ \texttt{mem["2:0x1068EEC"] = $m_0$}
  \end{footnotesize}
\end{center}

This way of simulating memory is not precise, because, for example, different symbolic values may point to the same memory. But we conjecture that in practice it shouldn't have a strong effect on the efficacy of the prediction model.

\subsection{Data Collection}

Algorithm \ref{alg:data-collection} describes our data collection process. It calls a few subroutines which we will not define as rigorously as Algorithm \ref{alg:data-collection}. These subroutines are fairly simple and depend heavily on third party tools and libraries:

\begin{itemize}
\item $\Call{extract\_functions}{exe}$ -  extracts functions from an executable $exe$ using radare2. Each function is represented as a pair $(address, opcodes)$, where $address$ is a function's virtual address and $opcode$ is the sequence of instructions the function consists of.

\item $\Call{get\_cfg}{f}$ - builds the CFG of a function $f$ using the recursive traversal algorithm \cite{schwarz-2002}. We use the NetworkX \cite{hagberg-2008-exploring} Python library to represent the function CFG. 

\item $\Call{get\_number\_of\_edges}{CFG}$ - returns the number of edges in a function's CFG.

\item $\Call{find\_longest\_path}{CFG}$ - finds the longest path through the function's CFG using NetworkX's method $all\_simple\_paths$. $all\_simple\_paths$ takes start and end vertices as arguments. 

\item $\Call{get\_path\_random\_walk}{CFG}$ - performs random walk through the CFG 30 times and returns the longest path.

\item $\Call{get\_op\_type}{opcode}$ - returns the radare2 type of an opcode. E.g. type of  the opcode \texttt{mov ebp, esp} is "mov".

\item $\Call{sym\_execute}{opcode}$  - executes the current opcode symbolically and updates the internal state of the symbolic execution engine.

\item $\Call{get\_call\_target}{opcode}$ - extracts the address of a call. E.g. for the call \texttt{call 0x01001C78} it  returns 0x01001C78.

\item $\Call{is\_api\_call}{address}$ - looks up an address in the IAT of the executable 

\item $\Call{look\_up\_number\_of\_arguments}{api\_name}$ - looks up the number of arguments an API function requires in a pre-constructed table.

\item $\Call{get\_arguments\_from\_stack}{n_{args}}$ - reads at most 12 arguments from the stack and removes $n_{args}$ values, where $n_{args}$ - is the actual number of arguments an API function takes and $3 \leq n_{args} \leq 12$.

\item $\Call{put\_in\_database}{api\_name, arguments, n_{args}}$ - puts the API name and its arguments from the stack, as well as the real number of arguments into the database.

\end{itemize}

\begin{algorithm}

\caption{Data Collection}
\label{alg:data-collection}
\begin{footnotesize}
\begin{algorithmic}
\State \textbf{Input:} $exe$ - 32-bit Windows executable
\State $functions \leftarrow \Call{extract\_functions}{exe}$
\ForAll{$f \in functions$}
	\State $CFG_f \leftarrow \Call{get\_cfg}{f}$
	\State $n_{edges} \leftarrow \Call{get\_number\_of\_edges}{CFG_f}$
	\If { $n_{edges} < 100$}
		\State $path \leftarrow \Call{find\_longest\_path}{CFG_f}$
	\Else
	\State $path \leftarrow \Call{get\_path\_random\_walk}{CFG_f} $
	\EndIf
	\ForAll{$opcode \in path$}
		\State $mnem \leftarrow \Call{get\_op\_type}{opcode}$
		\If {$mnem <> "call"$}
			\If {$mnem \in supported\_mnemonics$}
				\State $\Call{sym\_execute}{opcode}$ 
			\EndIf
		\Else
			\State $call\_target \leftarrow \Call{get\_call\_target}{opcode}$
			\If {$\Call{is\_api\_call}{call\_target}$}
				\State $name \leftarrow \Call{look\_up\_import\_table}{call\_target}$
					\State $n_{arg} \leftarrow \Call{look\_up\_number\_of\_arguments}{name}$
					\State $arguments \leftarrow \Call{get\_arguments\_from\_stack}{n_{args}}$
					\State $\Call{put\_in\_database}{name, arguments}$
			\EndIf
			\State $\Call{update\_eax}{ }$
		\EndIf
	\EndFor 
\EndFor
\end{algorithmic}
\end{footnotesize}
\end{algorithm}

We applied Algorithm \ref{alg:data-collection} to a data set of 2,185 32-bit PE files. The PE files are standard 32-bit Windows executables and DLLs that are typically found on a vanilla Windows 7 SP1 in C:\textbackslash{}Program Files, C:\textbackslash{}Windows and C:\textbackslash{}Windows\textbackslash{}system32. The only exception is a set Python 2.7 binaries. We skipped some of the files due to the large size which radare2 couldn't process efficiently at the time of the research.

From those binaries we extracted 63,195 calls with 2,748 unique API names. 
We used the top 25 API functions with the most data points (see Appendix \ref{app:api_list} for the full list of API names). Still the data was too biased with some API calls having thousands of entries and some just a few hundreds. To balance the dataset we sampled at most 400 calls from each group. The final dataset consists of N=9,451 samples of API calls and arguments read from the stack with each of the 25 API calls representing somewhere between 2.69\% and 4.91\% of the dataset.

\subsection{Argument Representation}
\label{sec:arg-repr}
Collected arguments can be one of the following types:
\begin{enumerate}
\item Integer - 0x1000, 0x80000002 etc.
\item Symbolic value - $eax, m_1,$ etc.
\item Symbolic expression - $edx + ebx - 1, 32 + ecx$, etc
\end{enumerate}

There are two problems.   First, the dataset is of mixed type.  Symbolic values and expressions are essentially strings, whereas the remaining arguments are integers.   Mixed types complicate the modeling process.\footnote{In particular, in this case, the emissions distributions of the HMMs would need to accommodate mixed types.  This would push the HMM out of the regime of standard HMMs and into the territory of hierarchical or mixture models, for which there may not be a closed form maximum likelihood solution for the E-step of Baum-Welch.}  Second, the vocabulary (i.e., the space of possible argument values) is too large.  The set of symbolic values and the set of symbolic expressions are both infinite and the set of integers [in 32-bit x86 CPU] has impractically large cardinality of $2^{32}$.   The problem here is that a model will be insufficiently powered to learn about the distribution of arguments if the number of arguments is too large relative to the number of samples.\footnote{In particular, in our case, the emissions distributions of the HMM are categorical with true multinomial parameter $p$ and estimator $\widehat{p}$.  The variance of $\widehat{p}$ grows monotonically with the size of the vocabulary, such that datasets with large vocabulary relative to sample size can be overfit and perform poorly in prediction mode.}  

To solve these problems, we map the three argument types into finite sets (with binning, as described below, for the integers).  This allows us to model arguments with a single categorical distribution of moderate size.   The mapping is performed as follows:

\textbf{Integer arguments} can be split into three groups:
\begin{enumerate}
\item \emph{Arbitrary values}, e.g. size of a memory allocation, or size of a buffer to read into. 
\item  \emph{Predefined values}, e.g. permission constants, flags, enumerations etc.
\item  \emph{Pointers}, i.e. addresses in the memory
\end{enumerate}

We conjecture that arbitrary values are not that important for API prediction. What might be important is their scale.  So, an arbitrary integer is encoded as the length of its string representation in hexadecimal (ignoring '0x' prefix). E.g. 0x1000 becomes 4, 0x12 becomes 2, etc. 

Predefined values on the other hand are crucial for prediction. For example, a system-defined registry key handle can be 0x80000001 (HKEY\_LOCAL\_USER), 0x80000002 (HKEY\_LOCAL\_MACHINE) etc., which makes it very likely that a function with one of  those values as a first argument is related to the registry. Therefore we do not abstract predefined values and leave them as is.

Finally, all the pointers are replaced by the string \emph{"ptr"} as a pointer can take any arbitrary value and depends on a compiler and compilation options.

\textbf{Symbolic values} are mapped onto the set of strings $\{reg, var, mem, ret, *\}$, where:
\begin{itemize}
\item $reg$ - corresponds to any register; since compilers can use some registers interchangeably we ignore the actual register names and sizes.
\item $var$ - corresponds to a function argument or a local variable. We make no distinction between arguments and local variables as any argument may or may not be put into a local variable before being passed to the callee.
\item $mem$ - is any symbolic memory value.
\item $ret$ - is any value that has been returned by a function.
\item $*$ - is a dummy value we use when the number of arguments is greater than the number of values on the stack. Since our symbolic execution engine supports only a small subset instructions, it is sometimes possible that the stack is in an inconsistent state.
\end{itemize}

\textbf{Symbolic expressions} can be any combination of symbolic values and supported operations between them, therefore the set of all potential values is infinite. Furthermore, it is very difficult to define a meaningful order relation on such set. So any symbolic expression is represented as the string \emph{"expr"}.

The vocabulary size from training (i.e. distinct arguments in our representation scheme) was $W=679$. Some samples of API calls  and their arguments are shown in Table~\ref{sample_sequences}.

\begin{table}
   \caption{Samples of API calls for 3 API call functions.}
  \begin{tabular}{ll}
  & \\
{\bf API Call Function} & {\bf Argument Sequence} \\
\hline
RegOpenKeyEx & var, var, 0x146, 1, 1  \\
RegOpenKeyEx & mem, 4, 0x170, var, 1\\
GetLocaleInfo & mem, 4, 1, 1\\
GetLocaleInfo & ret, 3, 2, 2\\
  SendDlgItemMessage &var, var, ret, expr, var, var, var, 1\\
  SendDlgItemMessage &mem, 0x411004, expr, 2, expr, 1, 1, expr\\
 
\end{tabular}
  \label{sample_sequences}
\end{table}

\subsection{Vectorization} \label{vectorization}

We conceptualize the arguments of an API call as a sequence, i.e. the ordering of the arguments matters.    To capture this sequence information, our approach applies what we call a \emph{Sequential (HMM)} vectorization to the dataset:

\begin{enumerate}
\item Train one or more relevant HMMs.\footnote{We fit HMMs efficiently using the Baum-Welch algorithm available in the \texttt{pyhsmm} Python package~\cite{johnson2013hdphsmm}. The package calls a forward-backward algorithm implementation which is written in C.} (What constitutes a ``relevant" HMM depends on the experiment; see the Experiment sections for more detail.)  
\item Feed a given argument sequence through the trained HMM(s) in order to represent it as a latent trajectory on the simplex.     
 \item Derive feature vectors from those latent trajectories. 
 \end{enumerate}
 
Many strategies could be used, in principle, for converting latent trajectories into features.  For this proof of concept, each API call's argument sequence is vectorized into $K+3$ features, where $K$ is the number of hidden states for the HMM.  The features are
\begin{itemize}
\item \emph{K hidden state features}: The log of the hidden state distribution at the terminal argument, i.e. the $K$-valued vector $\log P(s_T \cond y_{1:T}, \lambda)$
\item \emph{3 likelihood features}:   The log likelihood of the sequence, $\log P(y_{1:T} \cond \lambda)$, the mean log likelihood of an observation in the sequence, $\frac{1}{T} \log P(y_{1:T} \cond \lambda)$, and the log likelihood of the final observation, $\log P(y_T \cond \lambda)$.
\end{itemize}
Roughly, the hidden state features summarize the history of the argument sequence from the perspective of the final argument, and the the likelihood features summarize how unusual the sequence is relative to the relevant (training) population.  

\subsection{Prediction} \label{Prediction}

A predictive model, or classifier, is trained to learn a mapping between the feature vectors onto API function names.   Many classifiers could be used in principle; for this proof of concept, a multinomial logistic regression (MLR) is applied for simplicity.   The MLR produces, for each sample $i=1,\hdots, N$ represented as a feature vector $X_i$, a probability distribution over candidate API calls. If the candidate API calls are considered as a random variable $M$ and represented by indicators $m=1,\hdots, 25$, then the MLR model returns:
\[  p_{i,m} = P(M_i = m \cond X_i) \]
For the $i$th sample, the predicted API call $\widehat{M}_i$ is determined as the API call with the largest score
\[ \widehat{M}_i = \argmax_{m} \; p_{i,m} \]

To evaluate predictive performance, the dataset is randomly partitioned into a training set (7,869 samples, or 80\% of the dataset) and a test set (1,968 samples, or 20\% of the dataset).  

\section{Experiment 1}
\label{sec:experiment1}

For Experiment 1, we investigate the performance of the generic API call deobfuscator in an artificially constrained setting, where we treat each API call as having a known number of arguments.  (In Experiment 2, we relax this constraint.)  The experiment has two primary goals: (1) to investigate how well one can predict the unknown identity of the API call using our vectorization and prediction scheme;\footnote{ Note that, in reality, if we knew the number of arguments for an obfuscated API call, it would make the prediction task substantially easier. For example, if we knew that the obfuscated API call had 4 arguments, then we could immediately eliminate any candidate API functions that take a different number of arguments (e.g. \texttt{RegOpenKeyEx} and \texttt{SendDlgItemMessage} in Table~\ref{sample_sequences}).   However, we do not directly utilize this information here, as it would impede upon the primary goal of this experiment.} and (2) to investigate the extent to which the ordering of the arguments is important for these predictions.  

\subsection{Methodology}

For this experiment, we treat each API call as an i.i.d sample from a single population that includes all possible API functions.  Thus, we train a single HMM on the set of 7,869 API call samples that has no knowledge of the API function name.  For this experiment, we arbitrarily preset K=10.   We then use the \emph{Sequential (HMM)}  vectorization strategy, described in Section~\ref{vectorization}, to extract $K+3 = 13$ features from each of these samples based on its argument sequence.  Afterwards, we train a classifier, described in Section~\ref{Prediction}, to map feature vectors to API function names.  Predictions are made by applying the trained HMM vectorizer to the test set of 1,968 samples, and then running the resulting vectors through the trained classifier.  We take the API function with the highest score to be the predicted API function for that particular call.   We investigate the quality of this prediction.

A subsidiary purpose of this experiment is to address the question: how do we know that the sequence of arguments matters?   Might it be enough to represent the arguments as a \emph{bag of words}, whereby we consider the presence/absence of possible arguments but not their order?   To address this question, we apply an alternate vectorization strategy, which we refer to as the \emph{Bag of Words (SVD)} vectorization strategy. Here, we use one-hot encoding to create a $(N \times W)$ binary matrix such that the $(i,j)$th entry is set to 1 if the $i$th sample contains argument $j$, and 0 otherwise.   We then run a singular value decomposition (SVD) on this matrix, rotate it onto its right singular vectors, and use the $K+3$ dominant components to represent the dataset.    This strategy provides the best rank $K+3$ approximation to the original dataset, and allows for a comparison between the \emph{Sequential (HMM)} and \emph{Bag of Words (SVD)} vectorization schemes where the number of features is held constant. 

\subsection{Results}

The generic API call deobfuscator achieved 75.50\% accuracy in predicting the correct API call from a bank of 25 candidates when using our preferred argument sequence (HMM) vectorization strategy and a multinomial logistic regression classifier.   For comparison, random guessing would produce only 4.0\% accuracy, and random guessing by selecting the most common API call in the training set would produce only 4.9\% accuracy.   

\begin{figure*}
\centering
\begin{subfigure}{.48\textwidth}
  \centering
  \includegraphics[width=\linewidth]{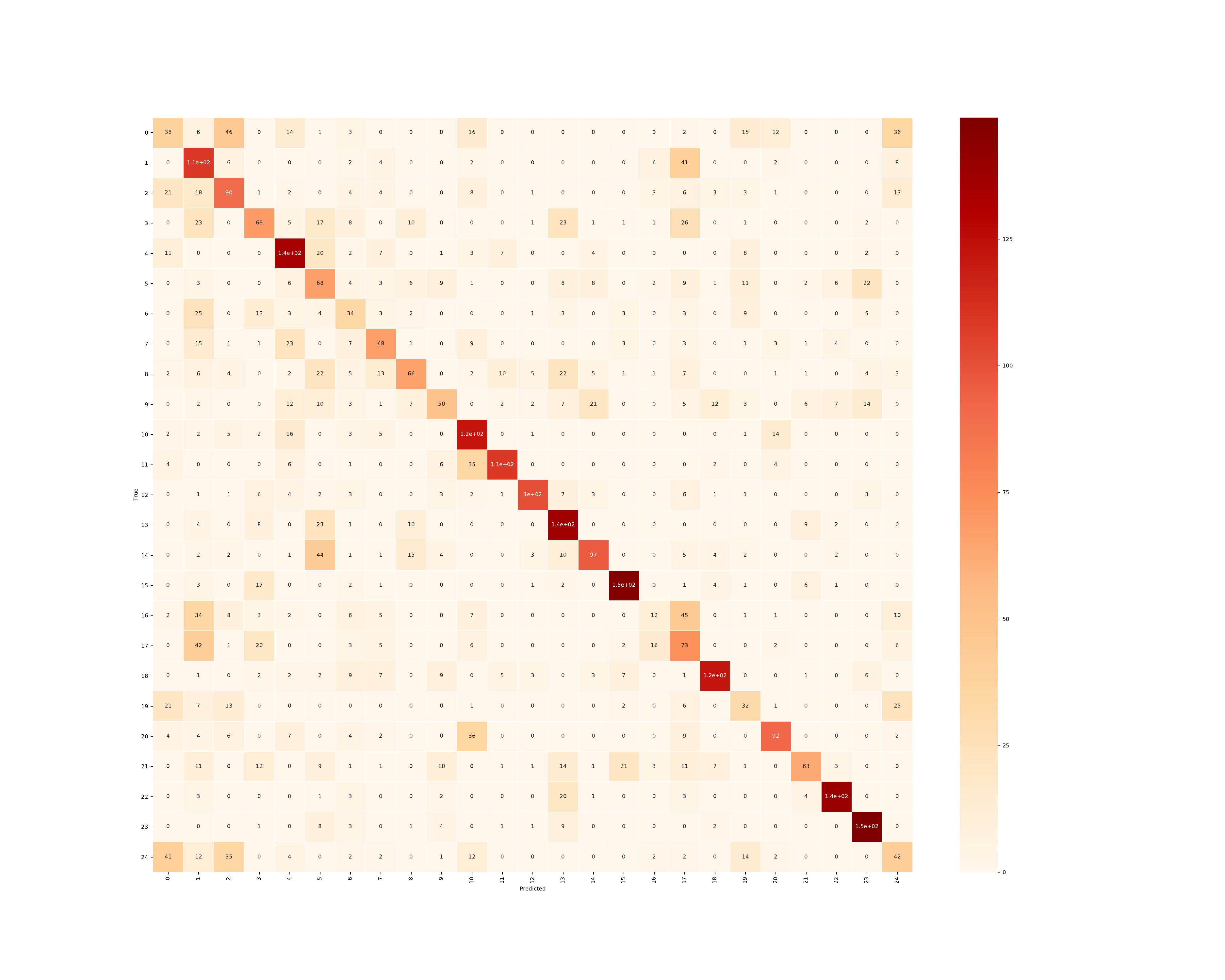}
 \caption{Bag of words (SVD) vectorization}
\label{a}
\end{subfigure} %
\begin{subfigure}{.48\textwidth}
  \centering
  \includegraphics[width=\linewidth]{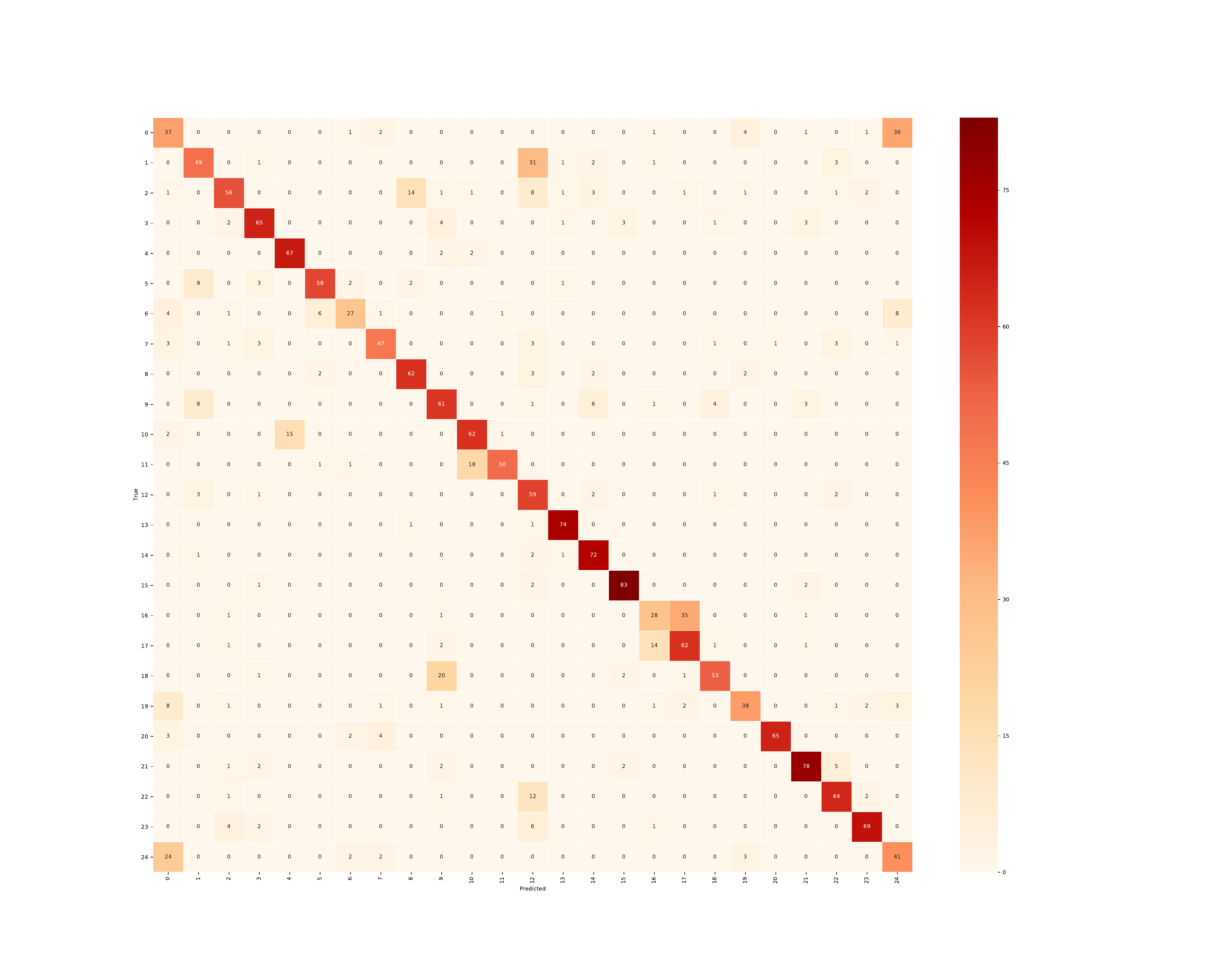}
\caption{Sequential (HMM) vectorization}
\label{b}
\end{subfigure}
\caption{\emph{Confusion matrices for API call classifier under different vectorization strategies.} The $(i,j)$th entry of the confusion matrix reports the number of times that the model predicted the  $i$th API call and the $j$th API call was the true call.  
A model is more accurate if it looks closer to a diagonal matrix (i.e. has higher counts -- looks more red -- on the diagonal and lower counts -- looks more beige -- on the off diagonal).    }\label{confusion_matrices}
\end{figure*}

 Predictive performance drops from 75.50\% under the sequential (HMM) vectorization strategy to 51.30\% under the bag of words (SVD) vectorization strategy.   As a result, the sequential (HMM) vectorization strategy causes the confusion matrix to concentrate its probably mass more tightly on the diagonal. (See Figure~\ref{confusion_matrices}).\footnote{Figure~\ref{confusion_matrices} also suggests that when the predictive model is wrong, the model tends to confuse the true API call with only a very small number of other calls.   Indeed, a follow up analysis revealed that the generic API call deobfuscator included the correct API call in its top prediction 75.50\% of the time, top 2 predictions 89.26\% of the time, and top 3 predictions 92.38\% of the time. } Thus, the \emph{sequence} (or ordering) of arguments appears to provide information valuable for predicting the API function name. 

\subsection{Discussion}

The major shortcoming of this experiment is that the  number of arguments for each API call is treated as known.  In practice, when we encounter an API function call whose name we would like to predict, we don't know how many arguments on the stack are intended for this call. The stack contains an arbitrary number of values, some of which remain from previous non-API calls, and some of which are pushed in advance for the subsequent calls (API or non-API).

\section{Experiment 2 }
\label{sec:experiment2}
For Experiment 2, we investigate the performance of the generic API call deobfuscator in a more realistic setting, where, due to obfuscation, we do not know the number of arguments that correspond to a given obfuscated API call.  Here, the model must evaluate numerous possible argument lengths based on the contents of the stack, thereby implicitly inferring the number of arguments.   Although this setting is more challenging, we also exploit it to our advantage.  In particular, as we consider popping different numbers of arguments from the stack, different subsets of API function names should become viable candidates.   This contrasts with Experiment 1 which implicitly ignored the known correspondence between API function names and argument number (e.g. {\tt RegOpenKey} takes 5 arguments, but {\tt GetLocaleInfo} takes 4 arguments) by fitting a single HMM to the entire corpus. 

\subsection{Methodology}

We imagine incrementally popping arguments from the stack to construct an argument sequence of increasing length $\ell$, and then evaluating the fit of that argument sequence against all API functions which are known to take $\ell$ arguments.   We then determine which API call is most likely, across all $\ell$, thereby also indirectly discovering the true $\ell^*$.  

To do this, we train $M=25$ separate HMMs, where each HMM is trained on all samples for a particular API function. \footnote{Note that we are now treating each API call sample as an i.i.d sample from the population of some \emph{particular} API function, rather than from a population of all API function, as in Experiment 1.}   For an arbitrary $m$th HMM model, we vectorize using the number of arguments $\ell_m$ appropriate for the corresponding API name.\footnote{If the candidate API call requires more arguments than can possibly be extracted from the stack, we set all the features to a default ``unlikely" value of -200.00.  Recall that all features are probabilities on a log scale, and so this unlikely value corresponds to a probability of $1.38\times 10^{-87}$.}   In this way, we obtain $ M*(K+3) = 325$ features for each sample.  As in Experiment 1, we arbitrarily fix the number of latent states for each HMM to $K=10$.


\subsection{Results}

\begin{figure}[h]
  \centering
  \includegraphics[width=\linewidth]{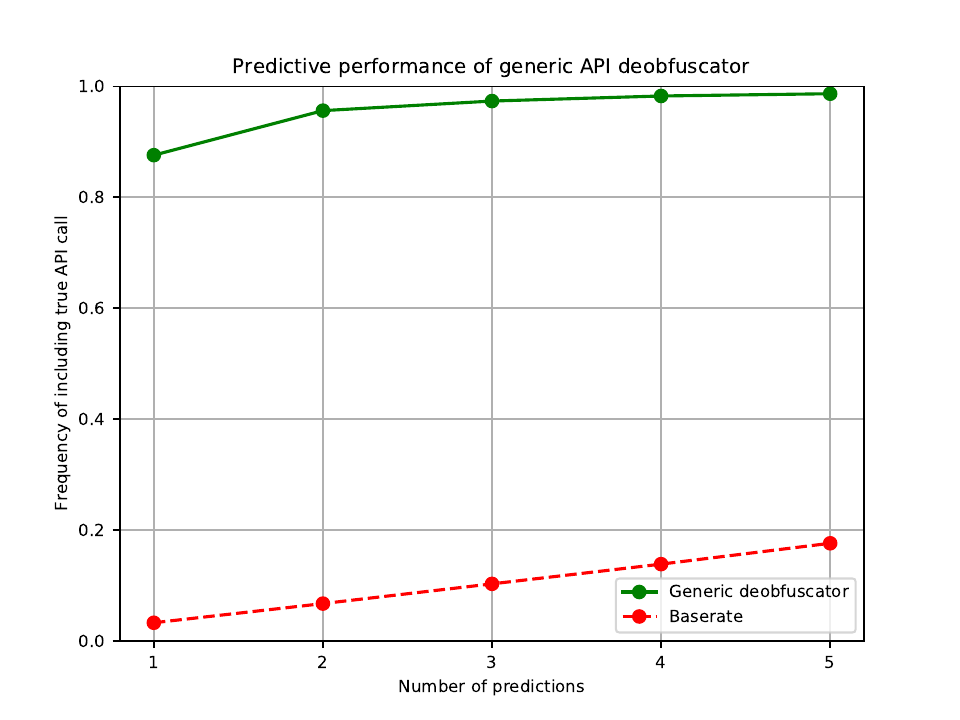}
\caption{\emph{Predictive performance of generic API call deobfuscator}.}
  \label{expt_2}
\end{figure}

Fig.~\ref{expt_2} shows the performance of the generic deobfuscator on a hold-out test set of obfuscation API calls, as a function of the number of predictions allotted to the model.   The generic API call deobfuscator selected the true API call 87.60\% of the time\footnote{Accuracy using only the $K$ latent state marginal features was 85.16\%, and using only the 3 likelihood features was 84.09\%.}  when it was allowed a single prediction and 95.63\%, 97.36\%, 98.27\%, and 98.68\% of the time when it was allowed 2, 3, 4, and 5 predictions, respectively.  In contrast, a baserate model, which guesses according to which API calls were most commonly observed during training, made correct predictions 3.25\%, 6.71\%, 10.27\%, 13.83\%, and 17.59\% of the time with those same prediction allotments.

\subsection{Discussion}
Although Experiment 2 handles relatively tougher constraints (the unknown number of arguments for an obfuscated API call) than Experiment 1, it actually yields substantially \emph{better} predictive performance.   This is because the methodology of Experiment 2 is more refined, exploiting the linkage between API function names and argument lengths to achieve a more informative vectorization.  
 
\section{Future Work}
\label{sec:future-work}

Now that we have evidence that our idea works in principle, we can start improving upon each subcomponent of our research: (1) the symbolic execution engine; (2) the data set; (3) the API prediction models.

\textbf{Improvements to the symbolic execution engine}:
\begin{itemize}

\item Extract additional contextual information about the API calls, such as sequences of API calls, data flow between the API calls, etc. This will provide additional information to the prediction model and potentially improve accuracy.

\item Support full x86 instruction set and test whether this improves the accuracy of prediction as well as cover all control flow paths with API calls.
\end{itemize}

\textbf{Improvements to the data set}
\begin{itemize}
\item Use a larger dataset of Windows executables, include popular programs and utilities other than those coming with Windows. Include malicious programs in the dataset. The malicious dataset must be as diverse as possible i.e. using various types of malicious functionality: code injection, command and control communication, encryption etc. 
\item Perform exploratory data analysis on malware dataset and prepare a list of API functions most likely to be seen in malware. Correlate API functions used with types of malicious behavior. Incorporate native APIs into the study as malware is known to use those too.
\end{itemize}

\textbf{Improvements to predictions model}:
\begin{itemize}
\item Exploit Bayesian HMM's with additional hierarchical structure.  For instance, apply hierarchical emissions distributions, which nests particular argument tokens within their broader categorizations (e.g. 0x146 and 0x170 are both \emph{flags}, and so observing one should increase the probability of the other).   Also, the 25 separate HMM's should be viewed as perturbations of a global HMM, so the learned emissions distribution for one API function should influence the estimated emissions distribution for the other API calls.
\item Optimize model parameters (e.g. $K$, the number of hidden states in the HMM, as well as parameters in the current classifier or alternative classifiers -- regularized MLR, SVM, etc.)
\item Seek richer vectorization strategies.  With HMM vectorization, we discard the latent state trajectories except for the terminal point; the rest of the trajectory implicitly enters into consideration via the likelihoods, but may provide additional information as well.  Alternatively, we might explore alternate tools for vectorizing sequences, e.g. LSTM.
\end{itemize}


\section{Conclusion}
\label{sec:conclusion}
Our proof of concept research suggest that (1) it is possible to predict the name of an API function from the arguments passed to it by the program and (2) machine learning, specifically HMM, is instrumental in the API prediction.  The results suggest future research directions. For example, we learned that some API calls are so similar that they might be mixed up by the prediction models. Using hierarchical HMMs and incorporating more contextual information (such as API sequences) might improve the accuracy.

\bibliography{references}{}
\bibliographystyle{ieeetr}


\appendices

\section{List of API Names}
\label{app:api_list}

\begin{enumerate}
\item CheckDlgButton
\item CoCreateInstance
\item CompareString
\item CreateFile
\item CreateWindowEx
\item DeviceIoControl
\item FormatMessage
\item GetLocaleInfo
\item HeapAlloc
\item LoadString
\item MessageBox
\item MultiByteToWideChar
\item PostMessage
\item ReadFile
\item RegCreateKeyEx
\item RegOpenKeyEx
\item RegQueryValueEx
\item RegSetValueEx
\item SendDlgItemMessage
\item SendMessage
\item SetDlgItemText
\item SetTimer
\item SetWindowPos
\item WideCharToMultiByte
\item WriteFile
\end{enumerate}

\end{document}